\documentclass[twocolumn,amsmath,amssymb,aps,prb,showpacs]{revtex4-1}
\usepackage{graphicx}

\begin{document}


\title{Hund's metallicity and orbital-selective Mott localization of CrO$_{2}$ in the paramagnetic state}
\author{Mu-Yong Choi}
\affiliation{Department of Physics, Myongji University, Yongin 17058,
Korea}

\begin{abstract}
We present electronic structure calculations of a CrO$_{2}$ compound in
the paramagnetic state within the computational scheme of
density-functional theory combined with dynamical mean-field theory. We
find that CrO$_{2}$ in the paramagnetic state is a strongly-correlated
Hund's metal with mixed-valent Cr ions. At high temperatures, CrO$_{2}$
shows an orbital-selective Mott phase, in which the Cr $d_{xy}$ orbital
is Mott-localized while the other Cr $t_{2g}$ $d$ orbitals remain
itinerant. When the temperature is lowered, the orbital-selective Mott
state is found to evolve into a Kondo-like heavy-fermion state with a
small pseudo gap at the Fermi level if the system remains in the
paramagnetic state.
\end{abstract}

\maketitle

\section{INTRODUCTION}
Chromium dioxide (CrO$_{2}$) is a ferromagnetic metal with a Curie
temperature of $\sim$390 K.~\cite{kouvel1967magnetic} Ferromagnetic
CrO$_{2}$ exhibits half-metallic behavior with one of the two spin bands
metallic and the other insulating.~\cite{schwarz1986cro2} Since fully
spin-polarized metals can be very useful in spintronics applications,
significant attention has been given to
CrO$_{2}$,~\cite{chamberland1977chemical} though the role of
electron-electron interactions and the correlated electronic structure
in CrO$_{2}$ have been embroiled in controversy and remains a subject of
active research.~\cite{[][{, and references
therein.}]katsnelson2008half,[][{, and references
therein.}]sperlich2013intrinsic}

Electron-electron correlation is a key concept in understanding
transition-metal oxides. A fascinating consequence of electron-electron
correlation is the Mott metal-insulator transition in partially-filled
bands, a proper explanation of which requires effort beyond the
traditional single-electron approach to solids. A consistent theoretical
framework for understanding of this phenomenon is provided via dynamical
mean-field theory (DMFT).~\cite{georges1996dynamical} DMFT studies show
that the Mott physics gets richer in multiband
systems.~\cite{anisimov2002orbital,koga2004orbital,liebsch2003mott,liebsch2007subband,werner2007high,de2009orbital,kita2012mott,jakobi2013orbital}
A particular example is the orbital-selective Mott transition in
partially filled multi-orbital systems, for which the Mott localization
takes place in some orbitals, while the rest remains
itinerant.~\cite{anisimov2002orbital,koga2004orbital,werner2007high,de2009orbital,kita2012mott,yu2013orbital,jakobi2013orbital}
The most obvious condition that can lead to an orbital-selective Mott
transition is that orbitals have different intraorbital Coulomb
repulsion.~\cite{wu2008theory} For systems having the same intraorbital
Coulomb repulsion for all orbitals, degenerate correlated bands of
different bandwidths~\cite{anisimov2002orbital,koga2004orbital} or the
crystal-field splitting of original degenerate
bands~\cite{werner2007high,de2009orbital} has been identified as a
prerequisite for the orbital-selective Mott transition. In either case,
a sufficiently strong Hund's coupling is essential to develop the
orbital-selective Mott phase (OSMP). The Hund's coupling or interorbital
exchange may promote a strong differentiation of the correlation
strength among the different orbitals which leads to the
orbital-selective Mott localization.The idea of selective localization
in the same subshell has recently received a great deal of
attention.~\cite{neupane2009observation,yi2013observation}

In most compounds, the ratio of the local direct Coulomb interaction to
the bandwidth determines the strength of the correlations, while the
Hund's coupling plays a subsidiary role. However, it has been
shown~\cite{werner2008spin,haule2009coherence,hansmann2010dichotomy,de2011janus,georges2012strong}
that in systems with multiple correlated bands crossing the Fermi level,
the crucial role for strong correlation can be played by the Hund's
coupling, rather than by the direct Coulomb interaction. The strength of
electronic correlation in these systems is much more sensitive to the
Hund's coupling than to the direct Coulomb interaction and therefore
materials which exhibit this are  often referred to as Hund's
metals.~\cite{georges2012strong,yin2011kinetic} Hund's metals in the
paramagnetic state exhibit very large resistivity and Curie-Weiss-like
magnetic susceptibility associated with well-formed local magnetic
moments. This picture is significant for transition-metal oxides with
partially-filled $t_{2g}$ bands that are well separated from the empty
$e_g$ band by a crystal field. Some of iron-based high-temperature
superconductors are a good example of Hund's
metals.~\cite{haule2009coherence,hardy2013evidence,lanata2013orbital}
When a crystal field further lowers the symmetry and induces additional
splitting in the original degenerate $t_{2g}$ bands, the Hund's coupling
may also develop the orbital-selective Mott localization in the
transition-metal oxides.

CrO$_{2}$ forms a rutile structure with space group $D_{4h}^{14}$:
$P4_{2}/mnm$. The unit cell contains two CrO$_{2}$ formula units. The
Bravais lattice is tetragonal with lattice constants $a=b=0.4421$ nm and
$c=0.2917$ nm.~\cite{porta1972chromium} The Cr atoms form a
body-centered tetragonal lattice and are surrounded by octahedra of O
atoms. The octahedra are slightly distorted away from the ideal
geometry, with the apical O atoms slightly more distant from the central
Cr atom than the equatorial O atoms. In the octahedral crystal field,
the Cr 3$d$ orbitals are split into a low-energy $t_{2g}$ triplet and a
high-energy $e_g$ doublet. Distortion of the Cr-O octahedron further
lifts the triple degeneracy of the $t_{2g}$ orbitals, leading to three
bands with predominantly $xy$ and $yz \pm zx$ characters. If Cr is in
its formal 4+ valence state, the remaining two $d$ electrons are
expected to occupy the $t_{2g}$ valence bands. Density-functional-theory
(DFT)
calculations~\cite{schwarz1986cro2,lewis1997band,mazin1999transport,toropova2005electronic,korotin1998cro}
show that the Cr $t_{2g}$ orbitals are strongly hybridized with the O
2$p$ orbitals to form dispersive bands crossing the Fermi level (EF).
The O 2$p$ bands also extend to EF and act as hole reservoirs, resulting
in Cr being mixed valent.~\cite{korotin1998cro} Measurements of the
local magnetic susceptibility in the paramagnetic phase reveal a
Curie-Weiss-like behavior with a local moment of
$\sim2\mu_B$.~\cite{chamberland1977chemical} Resistivity measurements
show that CrO$_{2}$ is a bad metal at high temperatures with the
resistivity which exceeds the Mott limit.~\cite{suzuki1998resistivity}
Both measurements demonstrate the possibility that the Cr $d$ electrons
are strongly correlated in the paramagnetic state. The local direct
Coulomb interaction ($U$) and Hund's coupling ($J$) between $d$
electrons, evaluated via constrain method, are as large as $U=3$ eV and
$J=0.87$ eV.~\cite{korotin1998cro} This indicates that both the Hund's
coupling and the direct Coulomb-interaction between $d$ electrons are
quite strong for CrO$_{2}$. These characteristics make CrO$_{2}$ a
strong candidate for a Hund's metal, with a possible OSMP in the
paramagnetic state.

The density-functional theory combined with the dynamical mean-field
theory (DFT+DMFT) is a powerful theoretical method for studying
strongly-correlated materials beyond the limit of the standard
density-functional theory.~\cite{kotliar2006electronic} The DFT+DMFT
calculations of CrO$_{2}$ in the ferromagnetic state have been performed
previously, showing the importance of dynamical correlation effects and
the Hund's coupling in the material without considering the
possibilities of Hund's metallicity and
OSMP.~\cite{craco2003orbital,chioncel2007half} In this paper, we present
the electronic structure calculations of the CrO$_{2}$ compound in the
paramagnetic state within the DFT+DMFT computational scheme, with a
focus on the possibilities of Hund's metallicity and OSMP.

\section{METHODS}
The charge-self-consistent DFT+DMFT calculations are performed as
presented in Ref.~\onlinecite{haule2010dynamical}. The DFT part of the
DFT+DMFT calculations is carried out employing the full-potential
linearized augmented-plane-wave band method implemented in the WIEN2k
package~\cite{blaha2001wien2k}. The generalized gradient approximation
with Perdew-Burke-Ernzerhof exchange-correlation
functional~\cite{perdew1996generalized} is adopted to describe the
exchange and correlation potentials. A $8\times8\times13$ $k$-point grid
is used for Brillouin-zone integrations. The continuous-time quantum
Monte Carlo method~\cite{haule2007quantum,gull2011continuous} is used as
an impurity solver of the DMFT part to yield the local self-energy of
the correlated $d$ electrons. We include the Slater form of the Coulomb
repulsion in its fully-rotationally-invariant
form~\cite{haule2010dynamical} to address the quantum impurity
problem.The Slater integrals $F^0$, $F^2$, and $F^4$ are determined
using the relations for $d$ orbitals, $U= F^0$, $J=(F^2+F^4)/14$, and
$F^4/F^2 = 0.625$.~\cite{himmetoglu2014hubbard} The maximum entropy
method~\cite{jarrell1996bayesian} is employed for analytical
continuation of the local self-energy to the region of real frequencies.

\begin{figure}
\includegraphics[width=8.5cm]{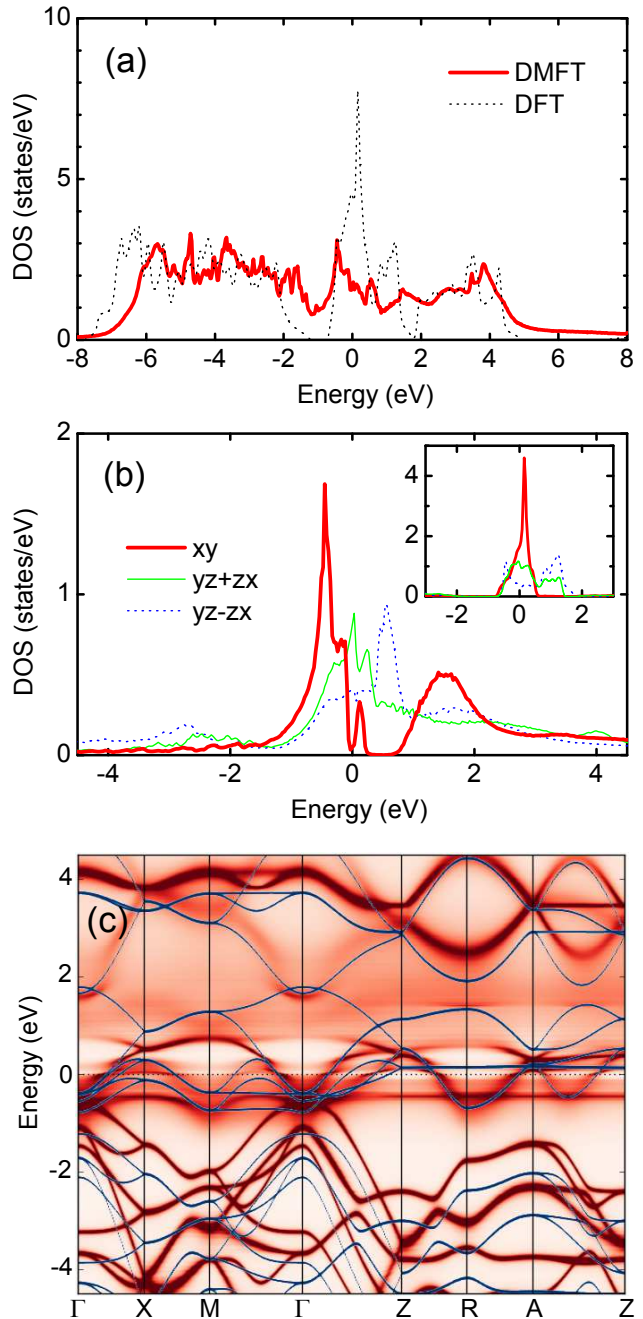}
\caption{\label{fig:1} (Color online) Band structure of CrO$_{2}$ in the
paramagnetic state at 400 K obtained from the DFT+DMFT calculations with
$U=3.0$ eV and $J=0.9$ eV. (a) Total density of states per formula unit,
compared to the results of the DFT calculations. The dotted line denotes
the corresponding DFT band structure. (b) Cr-$t_{2g}$ orbital-resolved
density of states. The inset shows the corresponding DFT band structure.
(c) Momentum-resolved total spectral functions along high-symmetry lines
in the Brillouin zone. The corresponding DFT bands are indicated as blue
lines.}
\end{figure}

\section{RESULTS AND DISCUSSION}
We find from the DFT+DMFT calculations that Cr in this material is mixed
valent, in agreement with the DFT calculations. The Cr valence is 2.4,
smaller than expected from the ionic picture. Figure~\ref{fig:1}
illustrates the band structure of CrO$_{2}$ at 400 K obtained from the
DFT+DMFT calculations. $U=3.0$ eV and $J=0.9$ eV are used in the
calculations, as suggested by the $ab$ $initio$
calculations~\cite{korotin1998cro,chioncel2007half} and previous
experimental works~\cite{sperlich2013intrinsic}. The total density of
states (total DOS) vs. energy plot in Fig.~\ref{fig:1}(a) demonstrates
that a DMFT treatment of electron correlations induces significant
modification in the band structure via DFT calculations. In comparison
with the DFT bands, the low-energy bands of predominantly O 2$p$
character below EF are pushed up by $\sim$0.6 eV and the quasi-particle
bands around EF, which are the Cr $t_{2g}$ bands hybridized with the O
2$p$ states, exhibit a reduced band width with the total-DOS peak
shifted below EF. The Cr-$t_{2g}$ orbital-resolved DOS of
Fig.~\ref{fig:1}(b) reveals that the total-DOS peak observed below EF
originates from the $xy$ orbital. At EF, the $xy$ band exhibits a pseudo
band gap and the $yz+zx$ band has a DOS peak. Neither the band gap nor
the DOS peak at EF can be identified in the DFT bands shown in the inset
of Fig.~\ref{fig:1}(b), which suggests that they originate from the
electron correlations. In Fig.~\ref{fig:1}(b), we also find broad local
maxima of DOS in the region between 1 and 3 eV, which are missing in the
DFT bands. The momentum-resolved total spectral functions along some
high-symmetry lines in Fig.~\ref{fig:1}(c) prove that the broad DOS
local maxima between 1 and 3 eV correspond to the upper Hubbard bands.
Localized in real space, Hubbard bands appear as a blurred region in the
momentum plot of spectral functions. The corresponding lower Hubbard
bands are less clear in the plot, because it has a very small spectral
weight distributed over a broad frequency region, as shown in
Fig.~\ref{fig:1}(b). The momentum plot of spectral functions also
reveals that the quasi-particle band near EF with prominent $d_{xy}$
character is nearly dispersionless, and is strongly correlated with a
large electron-scattering rate. The DFT+DMFT bands clearly indicate
strong correlations among $d$ electrons in this material.

\begin{figure}
\includegraphics[trim = 0mm 115mm 0mm 0mm, clip, width=8.5cm]{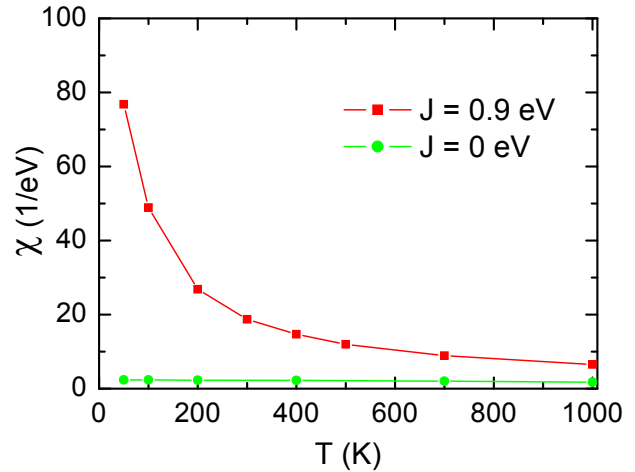}
\caption{\label{fig:2} (Color online) The local spin susceptibility of
$d$ electrons in a Cr atom as a function of temperature for different
Hund's couplings, $J=0$ and 0.9 eV.}
\end{figure}

\begin{figure}
\includegraphics[trim = 9mm 22mm 5mm 10mm, clip, width=8.5cm]{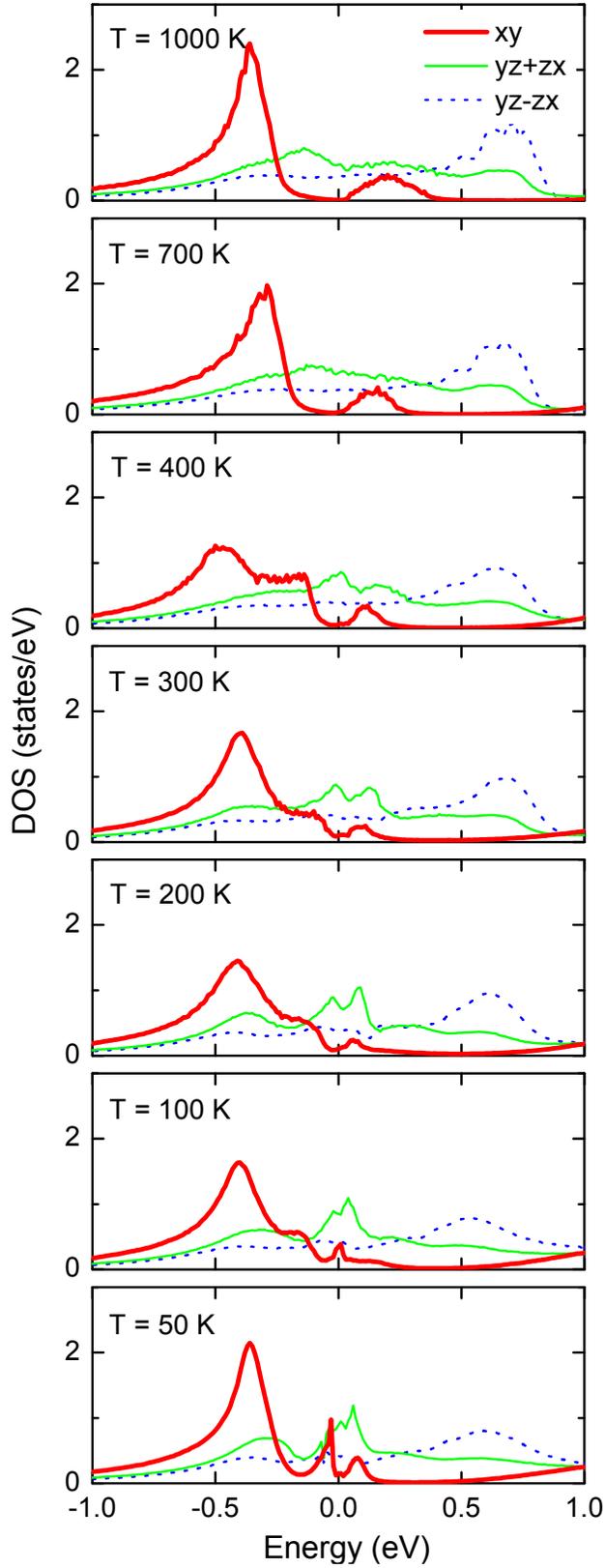}
\caption{\label{fig:3} (Color online) Evolution of the Cr-$t_{2g}$
orbital-resolved density of states with temperature.}
\end{figure}

One way of verifying the Hund metallicity is examining the dependence of
the magnetic susceptibility on the Hund's coupling strength as discussed
in Ref.~\onlinecite{haule2009coherence}. The local magnetic
susceptibility can be evaluated from the spin-spin correlation function:
\begin{equation}
{\chi_{loc}}=\int^{\beta}_{0}<S_z(0)S_z(\tau)>d\tau,
\end{equation}
where $S_z$ is the spin of the Cr atom and $\beta$ is the inverse
temperature.%
\footnote{In DMFT approximation, the local magnetic susceptibility is
the local spin susceptibility of the impurity atom, which is somewhat
different from the spin susceptibility of the lattice. Calculating the
lattice spin susceptibility requires much more elaborate efforts, as
discussed in Ref.~\onlinecite{georges1996dynamical}.} In
Fig.~\ref{fig:2}, we show the local spin susceptibilities of a Cr atom
both with and without the Hund's coupling. With the Hund's coupling
turned off, the susceptibility approximately obeys the Pauli law and the
DOS (not shown) resembles that of the DFT bands, indicative of
insignificant electron correlations in the absence of the Hund's
coupling. Doubling the direct Coulomb interaction brings marginal
changes in the susceptibility and DOS. In contrast, when the Hund's
coupling is turned on, the susceptibility becomes Curie-Weiss-like with
a local moment of $\sim1.5\mu_B$. The large local magnetic moment is a
hallmark of strong correlations. The magnetic-susceptibility
calculations indicate that with the Hund's coupling turned on, the
system becomes a strongly-correlated metal retaining the local nature of
magnetic moment. We thus conclude that CrO$_{2}$ is a Hund's metal, in
which the strength of electron-electron correlation is almost entirely
due to the Hund's coupling.

\begin{figure}
\includegraphics[trim = 9mm 22mm 5mm 5mm, clip, width=8.5cm]{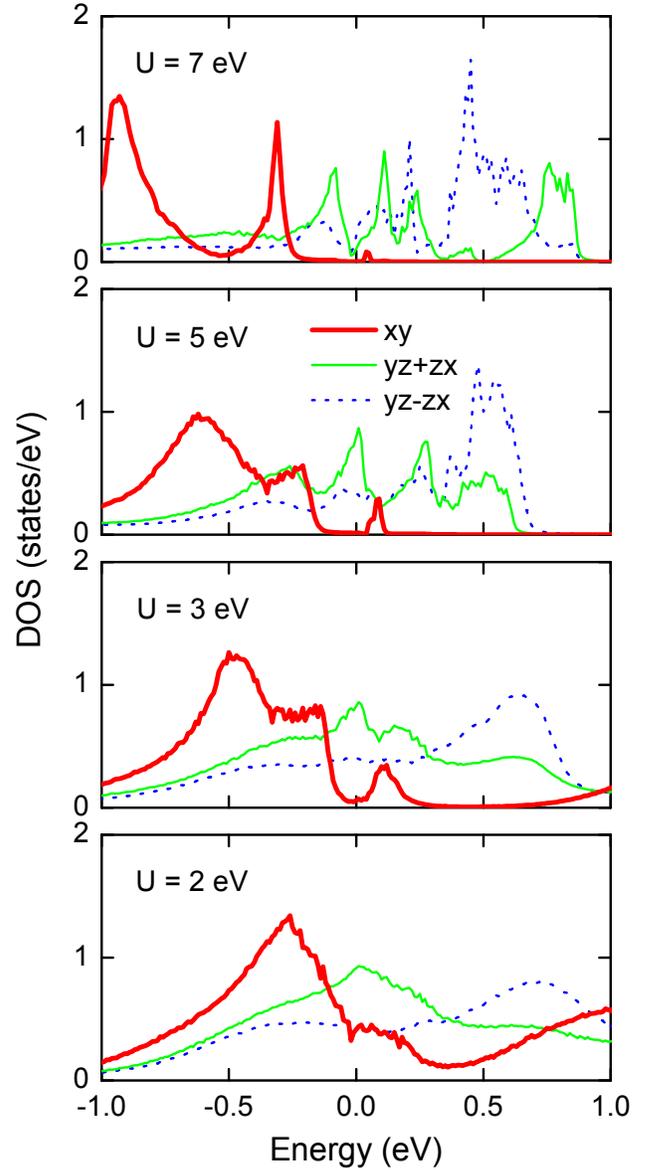}
\caption{\label{fig:4} (Color online) Cr-$t_{2g}$ orbital-resolved
density of states at 400 K for various $U$'s and the same $J/U= 0.3$.}
\end{figure}

We now focus our attention on the orbital-resolved DOS to investigate
the possibility of OSMP. Figure~\ref{fig:3} displays the evolution of
the Cr-$t_{2g}$ orbital-resolved DOS with temperature.%
\footnote{The orbital-resolved DOS at 400 K shown in Fig.~\ref{fig:3}
are slightly different from those in Fig.~\ref{fig:1}(b). The
discrepancy results from different frequency increments used for the
analytical continuation of the same self-energy to the region of real
frequencies. For detailed illustration of DOS near EF, a reduced
frequency increment is demanded as an input parameter for the analytical
continuation. Such an input-parameter-dependent outcome is an intrinsic
disadvantage of the maximum entropy method for analytical continuation.}
At high temperatures, the $xy$ orbital clearly exhibits a wide band gap
at EF and the $yz \pm zx$ orbitals exhibit a finite DOS at EF,
indicating that the strongly-correlated $xy$ orbital is in an insulating
state, while the other two orbitals are in itinerant states. The
insulating gap of the $xy$ orbital increases with enhanced $U$ above 3
eV and disappears when $U$ is lowered to 2 eV, as illustrated in
Fig.~\ref{fig:4}. These behaviors are consistent with the system forming
OSMP with the $xy$ orbital in the Mott insulating state. As the
temperature is lowered, the insulating gap of the $xy$ orbital
diminishes and is eventually superseded by a Kondo-like peak, as shown
in Fig.~\ref{fig:3}. For the metallic $yz+zx$ orbital, the Kondo-like
peak forms even at higher temperatures. When the temperature decreases
below 100 K, the Kondo-like peak of the $xy$ orbital splits to develop a
small pseudo gap at EF. It has been
shown~\cite{koga2005spin,de2005orbital,winograd2014hybridizing} that
hybridization between orbitals can replace OSMP with a Kondo-like
heavy-fermion regime at low temperatures. If the system actually enters
into the Kondo-like heavy-fermion regime at low temperatures, two states
are possible, either a heavy-fermionic metallic state or a Kondo
insulator with a small band gap between the bonding and antibonding
bands. We would thus say that CrO$_{2}$ in a high-temperature OSMP is
possibly driven by hybridization between $d$ orbitals to a
low-temperature Kondo-like metallic state which eventually acquires a
Kondo insulating gap below 100 K, if the system remains in the
paramagnetic state.

Most of the experimental studies on the electronic structure of
CrO$_{2}$ have been carried out on ferromagnetic samples. Recent
photoemission-spectroscopy measurements on ferromagnetic CrO$_{2}$
identifies the presence of the lower Hubbard band, indicating the
correlated Mott-Hubbard-type electronic structure of this
material.~\cite{sperlich2013intrinsic} Experimental work on the
electronic structure of paramagnetic CrO$_{2}$, to the best of our
knowledge, is limited to the temperature-dependent optical-conductivity
measurements across the ferromagnetic transition presented in
Ref.~\onlinecite{stewart2009ellipsometric}. The conductivity spectra
display a Drude contribution at low frequencies, a broad peak in the
mid-infrared region, and a sharp onset of interband absorption at
$\sim$1.5 eV demonstrating a distinct band gap. The main features of the
spectra appear to be consistent with the DFT calculations for the
ferromagnetic state showing EF lying in a band gap of the minority-spin
DOS and crossing the majority-spin
bands.~\cite{schwarz1986cro2,lewis1997band,mazin1999transport,toropova2005electronic,korotin1998cro}
However, the lack of temperature dependence of these features across the
transition is in conflict with the DFT calculations, which predict the
disappearance of the minority band gap and a drastically altered
spectral-weight distribution for the paramagnetic state. Within the DFT
scenario, the interband absorption above 1.5 eV is ascribed to interband
transitions across the minority band
gap.~\cite{stewart2009ellipsometric} The lack of temperature dependence
of the sharp interband absorption above 1.5 eV implies the presence of a
corresponding band gap on both sides of the ferromagnetic transition.
For the DFT+DMFT bands, it is tempting to assign the distinct absorption
region above 1.5 eV to transitions from the quasi-particle bands near EF
to the upper Hubbard bands which would stay put even below the Curie
temperature. The experimental confirmation of OSMP for this material has
yet to be made. Comprehensive photoemission-spectroscopy measurements
above the Curie temperature should be a great help.

\section{CONCLUSIONS}
To conclude, we find from DFT+DMFT calculations that CrO$_{2}$ in the
paramagnetic state is a strongly-correlated Hund's metal. At high
temperatures, CrO2 shows an orbital-selective Mott phase in which the Cr
$d_{xy}$ orbital is Mott localized, while the other $t_{2g}$ $d$
orbitals remain itinerant. With the temperature lowered, the
orbital-selective Mott state is found to evolve into a Kondo-like
heavy-fermion state with a small pseudo gap at EF if the system remains
in the paramagnetic state.



\bibliography{CrO2_resubmit}

\end{document}